\documentclass[aps,prb,a4paper,twocolumn,showpacs,floatfix,citeautoscript,superscriptaddress]{revtex4-1}
\usepackage{graphicx}
\usepackage{amsfonts}
\usepackage{amsmath}
\usepackage{amssymb}
\usepackage{bm}

\usepackage{bbold}
\usepackage{mathtools}
\usepackage{dcolumn}
\usepackage{hyperref}
\usepackage[usenames]{color}
\usepackage{xcolor}
\usepackage{subcaption}
\usepackage{cleveref}
\Crefformat{figure}{Fig. #2#1#3} 
\Crefmultiformat{figure}{Figs. ~#2#1#3}{ and ~#2#1#3}{, #2#1#3}{ and ~#2#1#3}
\Crefformat{section}{Sec. #2#1#3} 
\Crefformat{equation}{Eq. (#2#1#3)} 
\Crefmultiformat{equation}{Eqs. ~#2#1#3}{ and ~#2#1#3}{, #2#1#3}{ and ~#2#1#3}

\usepackage{physics}
\usepackage{units}
\usepackage{orcidlink}

\usepackage{ragged2e}
\DeclareCaptionJustification{justified}{\justifying}
\newcommand{\out}[1]{{}}

\newcommand{\fref}[1]{Fig.~\ref{#1}}

\def\vk{{\bf k}}

\begin{document}

\title{Relation between crystal structure and optical properties\\
in the correlated blue pigment YIn$_{1-x}$Mn$_x$O$_3$  
}
\author{Valentin Ransmayr}
\address{Institute of Solid State Physics, TU Wien, 1040 Vienna, Austria}
\author{Jan M.~Tomczak\,\orcidlink{0000-0003-1581-8799}}
\address{Institute of Solid State Physics, TU Wien, 1040 Vienna, Austria}
\author{Anna Galler\,\orcidlink{0000-0002-8596-7784}}
\address{Institute of Solid State Physics, TU Wien, 1040 Vienna, Austria}
\address{Max Planck Institute for the Structure and Dynamics of Matter,  22761 Hamburg, Germany}

\begin{abstract}
A material's properties and functionalities are determined by 
its chemical constituents and the atomic arrangement in which they crystallize.
For the recently discovered pigment YIn$_{1-x}$Mn$_x$O$_3$, for instance, it had been surmised that its bright blue color owes to an unusual, trigonal bipyramidal, oxygen coordination of the  manganese impurities. 
Here, we demonstrate that, indeed, a direct correspondence between details of the local Mn environment and the pigment's blue color holds:
Combining realistic many-body calculations (dynamical mean-field theory to treat the quasi-atomic Mn-multiplets at low doping $x=8\%$) with an effective medium description
(Kubelka-Munk model to describe scattering in a milled pigment sample), we find that only a Mn-coordination
polyhedra consisting of two distorted oxygen pyramids results in a diffuse
reflectance commensurate with the experimental blue color.
We motivate that the distortion of the bipyramid helps
circumventing atomic selection rules, allowing for dipolar $d$-$d$ transitions and creating the desired two-peak absorption profile.
\end{abstract}

\maketitle

\section{Introduction}
While some famous blue pigments---among them Ultramarin Blue (Lapis Lazuli), Han Blue and Egyptian Blue---have been known since antiquity, most synthetic blue pigment materials were discovered with the advent of modern chemistry. Some of them, such as cobalt blue CoAl$_2$O$_4$, Prussian blue Fe$_4$(Fe[CN]$_6$)$_3$ and azurite Cu$_3$(CO$_3$)$_2$(OH)$_2$~\cite{pigment_comp}, however, suffer from environmental or durability issues, such that the search for new earth-abundant and environmentally-benign alternatives has remained an active field of research. 
Thus, it was good news when a new, stable inorganic blue pigment material, YIn$_{1-x}$Mn$_x$O$_3$, was synthesized at Oregon State University in 2009.\cite{Subramanian2009} 
Discovered by serendipity, this novel blue pigment turned out to be ideal for many potential applications from ceramic glazes~\cite{Ocana2011,GOMES201811932,Gomes} and industrial coatings to plastics and artists' paint.\cite{shepherd_news}  It is being industrially 
produced 
under the name YInMn blue since 2017.\cite{shepherd_news,patent2012}

\begin{figure}[t!]
	\includegraphics[width=\columnwidth]{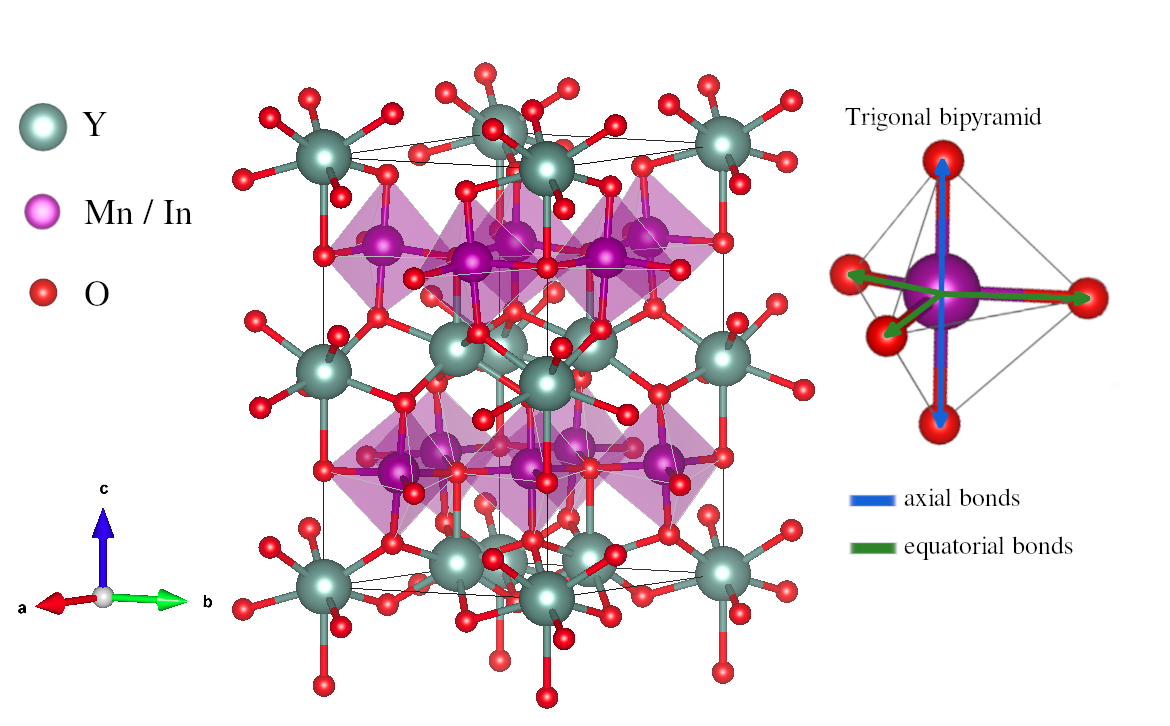}
	\caption{\textbf{Crystal structure of Y$M$O$_3$} (with $M$=Mn, In: purple spheres, O: red spheres, Y: green spheres). The structure consists of layers of corner-sharing, slightly tilted $M$O$_5$ trigonal bipyramids (TBP), separated by Y buffers. The inset on the right depicts a $M$O$_5$ TBP with equatorial and axial $M$-O bonds (visualization with VESTA~\cite{vesta}). }
	\label{fig:structure}
\end{figure}

YInO$_3$ and YMnO$_3$ are commonly known in the perovskite structure~\cite{Shannon1967}, but they can also be prepared in the hexagonal P6$_3$cm structure~\cite{Pistorius1976} shown in Fig.~\ref{fig:structure}: It 
consists of layers of corner-sharing, slightly tilted $M$O$_5$ ($M=\mathrm{In}$, Mn) trigonal bipyramids (TBP), separated by layers of Y$^{3+}$ ions. 
The researchers at Oregon State University were 
interested in potential multiferroic properties of hexagonal Y$M$O$_3$, when they noticed that replacing some In by Mn was inducing a brilliant blue color over a wide range of the paramagnetic YIn$_{1-x}$Mn$_x$O$_3$ (with $x=0.02-0.2$) solid-solution.~\cite{Subramanian2009} This finding was all the more surprising since pure YInO$_3$ and YMnO$_3$ are white and black, respectively.
From further experimental investigations of the materials' crystal and electronic structure, it was concluded that the blue color of YIn$_{1-x}$Mn$_x$O$_3$ must be a consequence of the unusual TBP coordination of the Mn$^{3+}$ ions, which determines the crystal-field splitting of the Mn 3$d$ shell.~\cite{Subramanian2009} A special role was attributed to the axial Mn-O bond distances, which are considerably shorter than the axial In-O bonds in YInO$_3$\cite{Subramanian2009,Amrani2012}. Indeed, further investigations~\cite{doi:10.1021/acs.chemmater.6b02827,Sarma2018} underlined the importance of the local crystalline environment around the Mn$^{3+}$ ions for the color of the pigment. In Ref.~\onlinecite{Sarma2018}, X-ray absorption near edge
structure (XANES) and extended X-ray absorption fine structure (EXAFS) spectroscopy investigations revealed the presence of two different TBP environments in YIn$_{1-x}$Mn$_x$O$_3$: one with symmetric and another one with asymmetric axial Mn-O bonds in the MnO$_5$ bipyramids. The distorted TBP, with asymmetric axial Mn-O bonds, was found to be dominant in the regime of low Mn concentrations $x$ and claimed to be responsible for the blue coloration. 
The experimental findings in Refs.~\onlinecite{Subramanian2009,Sarma2018} were supported by structural relaxations within density functional theory (LSDA+$U$) simulations.
However, no explicit calculation of the electronic structure and the optical response of YIn$_{1-x}$Mn$_x$O$_3$, which could provide a direct link between structural information and the optical transitions responsible for the blue color, has been presented yet. 

Indeed, systems like YIn$_{1-x}$Mn$_x$O$_3$ represent a challenge for modern electronic structure theory: First, a description of the Mn dopant and the correlation effects hosted by its open 3$d$ shell requires many-body methods beyond standard density functional theory (DFT-LDA). Second, the need for accurate optical gaps also calls for an advanced treatment of ligand states beyond simple DFT functionals.
Therefore, only few theoretical studies regarding the color of correlated pigment materials are available to date\cite{pnas_cesf,Galler2021_LnSF,Swagata_color}.
Blue pigments are particularly difficult to simulate, since they require a two-peak structure in the optical absorption which ensures that only blue light is reflected, while all other colors are absorbed. Red or yellow pigments, instead, are conceptually easier since they only require the absorption edge (hence the optical gap) to be at the right position.

In this work, we employ the recently developed mBJ@DFT+DMFT approach\cite{Galler2021_LnSF,Boust2021_mBJ,Galler2022_nitrides} to compute the electronic structure and optical conductivity of YIn$_{1-x}$Mn$_x$O$_3$. We focus on a Mn concentration of $x=8\%$, for which a brilliant blue color is observed in experiment. The methodology combines an improved description of non-local exchange-driven band-gaps via the semilocal modified Becke-Johnson (mBJ)\cite{tran_mbj_original,koller_mbj_2011} potential with an advanced treatment of the strongly correlated Mn 3$d$ states within dynamical mean-field theory (DMFT)\cite{Vollhardt_DMFT,Georges_DMFT}. This approach has already been successfully employed to compute the electronic structure and optical absorption edge in the rare-earth fluorosulfide pigments $Ln$SF (with $Ln=$ rare-earth).\cite{Galler2021_LnSF} We demonstrate that the same computational approach can be applied to YIn$_{1-x}$Mn$_x$O$_3$, where it gives valuable insight into the mechanisms that are responsible for the blue coloration of the pigment.
We further model the diffuse reflectance of the milled pigment using the effective medium description by Kubelka and Munk\cite{kubelkaMunk1931,Kubelka1948}, and compare our results to available experimental data.

\section{Computational approach} 
We start from the experimental P6$_3$cm crystal structure of YInO$_3$. We double the unit cell in $\vu*{c}$-direction and replace one of the 12 In atoms with Mn, which yields a Mn concentration of $x\approx8\%$ in the supercell. For numerical feasibility, we perform all calculations with this supercell, thus neglecting effects of disorder.
The $a$ and $c$ lattice parameters, which show a clear linear dependence with Mn concentration $x$, are extracted from experimental measurements of Ref.~\onlinecite{Sarma2018}. For $x\approx8\%$, this yields $a=\unit[6.2620]{\AA}$ and $c=\unit[12.1788]{\AA}$. We relax all internal coordinates within DFT, employing a full-potential linear-augmented plane-wave basis set, as implemented in the WIEN2k~\cite{wien2k,wien2k2020} program package. In the structural relaxations, 
a multisecant approach~\cite{wien2k2020}
with a $\vk$-grid of 100 $\vk$-points in the reducible Brillouin zone was used and atomic positions were relaxed until forces were less than 1 mRy/bohr. We performed relaxations, both, within LDA and LSDA$+U$. We found the LSDA+$U$-relaxed structure to be virtually unaffected by changes of $U$ within a window of $\unit[3-8]{eV}$. Also the type of the imposed (artificial) magnetic order (ferro- or antiferromagnetic) 
had no influence on the result.
The LDA-relaxed structure, instead, displayed a slightly different local crystalline environment around the Mn$^{3+}$ ions, which compared better to experiment\cite{Sarma2018} (see Results and the Appendices). We therefore focus on the LDA-relaxed structure, but compare the optical conductivity of both relaxations to highlight the
influence of the local crystal structure onto the color.   

To compute the electronic structure and optical conductivity of paramagnetic YIn$_{1-x}$Mn$_x$O$_3$ (with $x\approx8\%$), we employ the recently developed mBJ@DFT+DMFT approach, 
for details of the method see Ref.~\onlinecite{Boust2021_mBJ}. It basically consists of three steps:

(1) We start by performing a charge-self-consistent DFT+DMFT calculation~\cite{wien2k,Aichhorn2009,Aichhorn2011, triqs_dfttools}, in which we treat local correlations in the Mn 3$d$ shell with the quasi-atomic Hubbard-I~\cite{hubbard_1} DMFT solver.  
To this end, we construct projective Wannier functions~\cite{Aichhorn2009,triqs_dfttools} representing the correlated Mn 3$d$ states from all Kohn-Sham eigenstates enclosed by the energy window $[-9,10]$~eV  
around the Fermi level.
In the spirit of previous DMFT calculations for other transition-metal oxides~\cite{Leonid2021_octu,Aichhorn2020}, 
we mimic a $d$-only (downfolded) setting by limiting the hybridization function to
an energy window of $[-1,3]$~eV, accounting for 92\% of the Mn spectral weight from the large energy window.
We use the electronic structure code WIEN2k~\cite{wien2k} for DFT and the TRIQS~\cite{triqs_cpc} and TRIQS/DFTTools~\cite{triqs_dfttools} packages for DMFT.

(2) After converging the charge-self-consistent DFT+DMFT calculations, we run an additional DFT cycle employing the Tran-Blaha modified Becke-Johnson (mBJ) potential~\cite{tran_mbj_original,koller_mbj_2011}, as implemented in WIEN2k~\cite{wien2k}. Such a perturbative use of the mBJ potential  was shown to yield reliable values for semiconducting band gaps~\cite{hong_mbj_2013}.  For the problem at hand, we demonstrate (see Appendix~\ref{app1}) that a perturbative use of the mBJ potential yields a value of 3.71 eV for the band gap in pure YInO$_3$, which is in good agreement with the experimental value of around \unit[3.8]{eV}, estimated from the diffuse reflectance measured in Ref.~\onlinecite{Subramanian2009}.

(3) After the perturbative mBJ step, we finally recalculate the  electronic structure of YIn$_{1-x}$Mn$_x$O$_3$ by performing a DMFT cycle using the Hubbard-I approximation for the mBJ-corrected Kohn-Sham bands.

Our calculations do not break spin-symmetry (paramagnetic) and the Mn 3$d$ quantum impurity problem contains all five Mn 3$d$ orbitals.
The fully rotationally-invariant local Coulomb interaction on the Mn 3$d$ shell was parametrized by $U=\unit[2.5]{eV}$ and $J=\unit[0.8]{eV}$. 
Previous studies\cite{Subramanian2009} suggested the Mn-multiplet structure to be a crucial ingredient for the material's color. The Hubbard $U$ parameter was therefore deliberately chosen to be of a magnitude comparable to visible light. Ideally, the interactions should be computed from first principles in the future.
We further employed the fully-localized-limit double-counting correction in the atomic limit~\cite{Pourovskii2007}, i.e.\ $E_{DC} = U(N - 0.5) - J (0.5 N - 0.5)$ with the Mn$^{3+}$ atomic occupancy $N=4$.  All calculations were carried out at a temperature of \unit[290]{K}.

 As a result of the mBJ@DFT+DMFT calculations, we obtain the many-body spectral function $A_\mathbf{k}(\omega)$ encoding the excitation energies of an electron addition/removal into the many-body ground state.  
 $A_\mathbf{k}(\omega)$ is a crucial ingredient for determining the absorption properties of a crystalline bulk material.  By using linear response theory, and neglecting vertex corrections, the real part of the  frequency-dependent optical conductivity reads\cite{Dresselhaus_opt,PhysRevB.80.085117}
\begin{align}
\label{eq:cond}
\sigma_{\alpha\alpha}(\Omega)  &=\frac{2\pi e^2 \hbar}{V}  \sum_{k}  \int d\omega \;\frac{f(\omega-\Omega/2)-f(\omega+\Omega/2)}{\Omega} \nonumber \\
&\times \text{Tr}\left\{\mathbf{A_{\vb{k}}}(\omega-\Omega/2)\mathbf{v}_{\vb{k},\alpha}\mathbf{A}_{\vb{k}}(\omega+\Omega/2)\mathbf{v}_{\vb{k},\alpha}\right\} , \; 
\end{align}
where $V$ is the unit-cell volume, $\Omega$ the frequency of the incident light and $f(\omega\pm\Omega/2)$ Fermi functions which ensure that transitions take place only between occupied and empty states. As usual, we limit the calculation to direct optical transitions, without any momentum transfer.\cite{Dresselhaus_opt} $\mathbf{v}_{k,\alpha}$ are matrix elements of the momentum operator in the Cartesian directions $\alpha=x, y$ or $z$,\cite{AmbroschDraxl20061} and  $\mathbf{A}_\mathbf{k}(\omega)$ are the $\vk$-resolved spectral-function matrices in orbital space. 
The complex conductivity $\tilde{\sigma}(\omega)$ can be constructed from $\sigma(\omega)$ using a Kramers-Kronig transform.

To study the influence of individual orbitals and matrix elements in the optical response, we compare $\sigma(\omega)$ to the partial joint density of states
\begin{eqnarray}
    D_l(\Omega)&\propto& \int d\omega \;\frac{f(\omega-\Omega/2)-f(\omega+\Omega/2)}{\Omega} \nonumber \\
    &&\times A_l(\omega-\Omega/2)A_l(\omega+\Omega/2)
    \label{eq:JDOS}
\end{eqnarray}
where $A_l(\omega)$ is the local spectral function traced over the orbital character $l$.

When computing optical properties, we use a refined mesh of 1000 $\vk$-points in the reducible Brillouin zone, a frequency spacing of \unit[1]{meV} and an additional broadening of excitations $\delta=\unit[0.8]{meV}$.

Since pigments are used in powdered form or lacquers, i.e.\ as small particles within a transparent glaze, in principle a complicated multiple-scattering problem needs to be solved to obtain the diffuse reflectance $R$. 
A simple and commonly used shortcut is the effective medium description by Kubelka and Munk (KM)~\cite{kubelkaMunk1931,Kubelka1948}, which models the propagation of light through a homogeneous layer with a pigment concentration $c_\%$, that absorbs light with an amplitude $c_\%K(\omega)$ and backscatters it with rate $\beta(\omega)$. 
Here, $K(\omega)$ is the macroscopic absorption coefficient of the pigment's crystalline bulk, linked to the complex conductivity $\tilde{\sigma}$
via $K(\omega) = {\omega}/{c} \Im \left( \sqrt{1+\frac{4\pi i}{\omega} \tilde{\sigma}(\omega)} \right)$ with $c$ the speed of light.
The parameter $\beta(\omega)$---an inverse scattering length---contains information on the imperfect microscopic structure of the sample and is often treated as phenomenological and static ($\beta(\omega)\equiv\beta$).\cite{pnas_cesf}
In the KM model, the diffuse reflectance of a semi-infinite pigment layer is given by 
\begin{equation}\label{eq:rdiff}
\begin{split}
	R_{\infty}(\omega) &= \alpha - \sqrt{\alpha^2 - 1 } \; \text{, with}\\
	\alpha &= 1 + \frac{2c_\%K(\omega)}{\beta} \text{.}
\end{split}
\end{equation}
 Supplementing the simulated diffuse reflectance with the spectral distribution of a light source---we use the CIE standard illuminant D65\cite{D65} corresponding to daylight on Earth---and the empirical sensitivities for the color perception of the human eye\cite{xyz},
we compute the coordinates in the sRGB color space~\cite{srgb}, predicting the color of YIn$_{1-x}$Mn$_x$O$_3$. 

\begin{figure*}[t]
    \hspace{-1cm}
	\includegraphics[width=1.\textwidth]{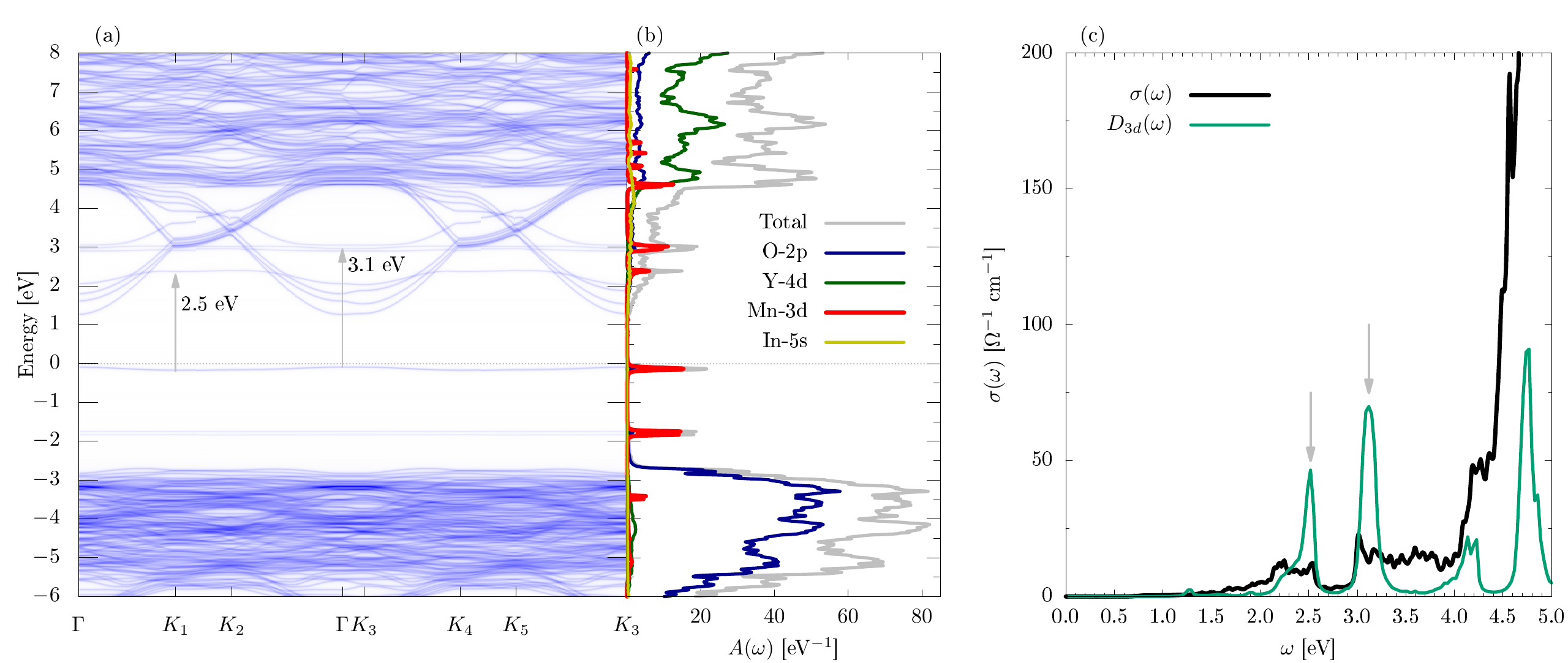}
	\caption{\textbf{Electronic structure and optical conductivity of YIn$_{0.92}$Mn$_{0.08}$O$_3$.} \textbf{(a)} $\vb{k}$-resolved and \textbf{(b)} $\vb{k}$-summed spectral function obtained from mBJ@DFT+DMFT. The valence band (below \unit[-2.8]{eV}) is formed by the O 2$p$ (blue) states. The conduction band involves some dispersive bands (starting a \unit[1.2]{eV}) with mainly In 5$s$ (yellow) character and Y 4$d$ (green) states above \unit[4.5]{eV}. The Mn 3$d$ states are split into multiplets by the on-site Coulomb interaction. The occupied Mn 3$d$ orbitals form impurity states within the band gap. In \textbf{(c)} we show the computed (polarization-averaged) optical conductivity $\sigma(\omega)$ (black). The two-peak structure between \unit[1.2-4]{eV}, which is crucial for the blue color, arises mainly from Mn $d-d$ optical transitions, as can be seen by comparing to the partial joint density of states $D_{3d}(\omega)$ (green). 
	}
	\label{fig:bs_dos_opt}
\end{figure*}

\section{Results}
\subsection{Crystal structure}
Our structural relaxations confirm the previously observed trend~\cite{doi:10.1021/acs.chemmater.6b02827,Sarma2018,Subramanian2009,Amrani2012} 
of decreasing axial $M$-O bond distances in the trigonal bipyramids (TBPs) of Y$M$O$_3$, when In$^{3+}$ ions are replaced with the smaller Mn$^{3+}$. The equatorial bond lengths are  less affected by the Mn substitution. Table~\ref{tab:MnO_bl_x8} further shows an axial distortion of the TBPs for small Mn concentrations, $x=8\%$,  expanding one of the axial Mn-O bonds by $\sim10\%$. These asymmetric  
bond lengths have indeed been observed in XANES and EXAFS experiments\cite{Sarma2018} and were claimed to be important for the blue coloration of YIn$_{1-x}$Mn$_x$O$_3$ as they change the local crystalline environment of the Mn$^{3+}$ ions. In our calculations, clearly asymmetric axial bond distances are observed only in the LDA-relaxed structure, see Table~\ref{tab:MnO_bl_x8}. The effect is much less pronounced for relaxations with LSDA+U (see Table~\ref{tab:large} in Appendix~\ref{app2}). 
For the electronic structure and optical calculations, we therefore proceed with the LDA-relaxed structure. For comparison, we also perform computations of the optical conductivity with symmetric axial Mn-O bond distances (LSDA+U relaxed structure). This way, we can analyse how the optical response and color of the pigment are affected by different local crystalline environments around the Mn$^{3+}$ ions, which have been argued\cite{Sarma2018} to be present in the disordered solid-solution.

\begin{table}
	\bgroup
	\def\arraystretch{1.5}
	\centering
	\begin{tabular}{c||c|c||c|c}
		$\ell$ [\AA]
		& \multicolumn{2}{c||}
		{YInO$_3$} & \multicolumn{2}{c}{YIn$_{1-x}$Mn$_x$O$_3$}\\
		\hline
		bond & 
		Exp.  & 
		LDA & 
		Exp. & LDA\\ 
		\hline
		\hline
		
		axial 1 & 2.089 & 2.088 & 1.879 & 1.891   \\
		\hline
		axial 2 & 2.093 & 2.111 & 2.087 & 2.045  \\
		\hline
		equatorial 1 & 2.097 & 2.130 & 1.976 & 1.862 \\
		\hline
		equatorial 2  & 2.127 & 2.122 & 1.976 & 1.862 \\
		\hline
		equatorial 3 & 2.127 & 2.122 & 2.087 & 1.885 \\
		\hline
	\end{tabular}
	\caption{$M$–O bond lengths $\ell$ of pure YInO$_3$ and YIn$_{1-x}$Mn$_x$O$_3$. Theoretical bond lengths were obtained using the LDA potential.
	Experimental data are reproduced from Table 2 of Ref.~\onlinecite{Sarma2018}. Note that our calculation for YIn$_{1-x}$Mn$_x$O$_3$ were performed for x=$8.33\%$, while the experiment~\cite{Sarma2018} was carried out for x=$5\%$.} \label{tab:MnO_bl_x8}
	\egroup
\end{table}

\subsection{Electronic structure and optical response}
We begin with the electronic structure of
crystalline bulk YIn$_{1-x}$Mn$_x$O$_3$ (with $x=8.33\%$).
\Cref{fig:bs_dos_opt}a shows the $\vb{k}$-resolved spectral function $A_{\vb{k}}(\omega)$, as obtained from mBJ@DFT+DMFT. The chosen $\vb{k}$-path through the Brillouin zone is specified in \Cref{fig:bz_x8} and the   $\vb{k}$-integrated spectral function $A(\omega)$ is depicted in \Cref{fig:bs_dos_opt}b.  The valence band below \unit[-2.8]{eV} (energies measured from Fermi level), 
is mainly formed by the filled O 2$p$ states. The conduction band starts at around \unit[1.2]{eV} with some dispersive bands of mainly In 5$s$ character, before Y 4$d$ bands appear above \unit[4.5]{eV}. The simulated O-2$p$ -- In-5$s$ band gap in YIn$_{0.92}$Mn$_{0.08}$O$_3$ thus amounts to \unit[4]{eV}, which is close to the predicted and experimental value for pure YInO$_3$ (see Appendix~\ref{app1}). 
At a concentration of $8\%$, the Mn atoms are fairly isolated.
Accordingly, their 3$d$ states form sharp peaks in the local spectral function, corresponding to  
only weakly-dispersive quasi-atomic multiplets. The most prominent part of the lower Hubbard bands, i.e.\ the occupied Mn 3$d$ states directly below the Fermi energy and at \unit[-1.8]{eV}, are located within the O-2$p$ -- In-5$s$  band gap. 
The multiplet peaks of the Mn-$3d$ upper Hubbard band lie within the dispersive conduction bands, with sizable features visible in a window of \unit[2-6]{eV}.

The Mn 3$d$ states play a crucial role in the optical response of YIn$_{0.92}$Mn$_{0.08}$O$_3$,  as can be seen in \Cref{fig:bs_dos_opt}c, which depicts the simulated optical conductivity $\sigma(\omega) \equiv \sum_{\alpha}\sigma_{\alpha \alpha}(\omega)/3$ averaged over polarizations $\alpha$
(see \Cref{eq:cond}) as well as the partial joint density of states (JDOS; see \Cref{eq:JDOS}) of the Mn 3$d$ states $D_{3d}(\omega)$.   
One can clearly see that $\sigma(\omega)$ exhibits two dominant peaks in the energy range from \unit[1.2-4]{eV}. Such a two-peak structure is crucial for any blue coloration since it allows for absorption on the low- as well as high-energy side of the optical spectrum, while reflecting the blue components (\unit[2.5-2.8]{eV}). 
The shallow onset of the first absorption peak at around \unit[1.2]{eV} stems from optical transitions from the highest occupied Mn 3$d$ states (red peak in $A(\omega)$ just below the Fermi energy) to the bottom of the dispersive conduction band. The principle weight of the absorption peak at around \unit[2-2.5]{eV}, instead, can be traced to $d-d$ transitions.
Indeed, a comparison with the partial JDOS $D_{3d}(\omega)$ suggests 
optical transitions from the highest occupied to the lowest unoccupied Mn 3$d$ states to be active.
In an atom, such $d-d$ optical transitions would be forbidden by the dipole selection rule $\Delta l=\pm 1$, with $l=d$. In a solid, the crystal field, hybridization with ligands, as well as non-local transitions can relax the optical selection rules so that $d-d$ processes become possible. 
The onset of the second prominent absorption peak at \unit[3]{eV} is in fact also dominated by optical $d-d$ transitions. In this case, the relevant transitions happen between the occupied Mn 3$d$ states just below the Fermi level and the unoccupied Mn 3$d$ states around \unit[3]{eV}. This second, broad peak in $\sigma(\omega)$ then continues till around \unit[4]{eV} with transitions from Mn 3$d$ into the conduction band. At around \unit[4]{eV}, optical transitions from the O 2$p$ valence band to the conduction band set in, leading to a rapid increase of the optical conductivity.

\begin{figure}[t]
	\includegraphics[width=\columnwidth]{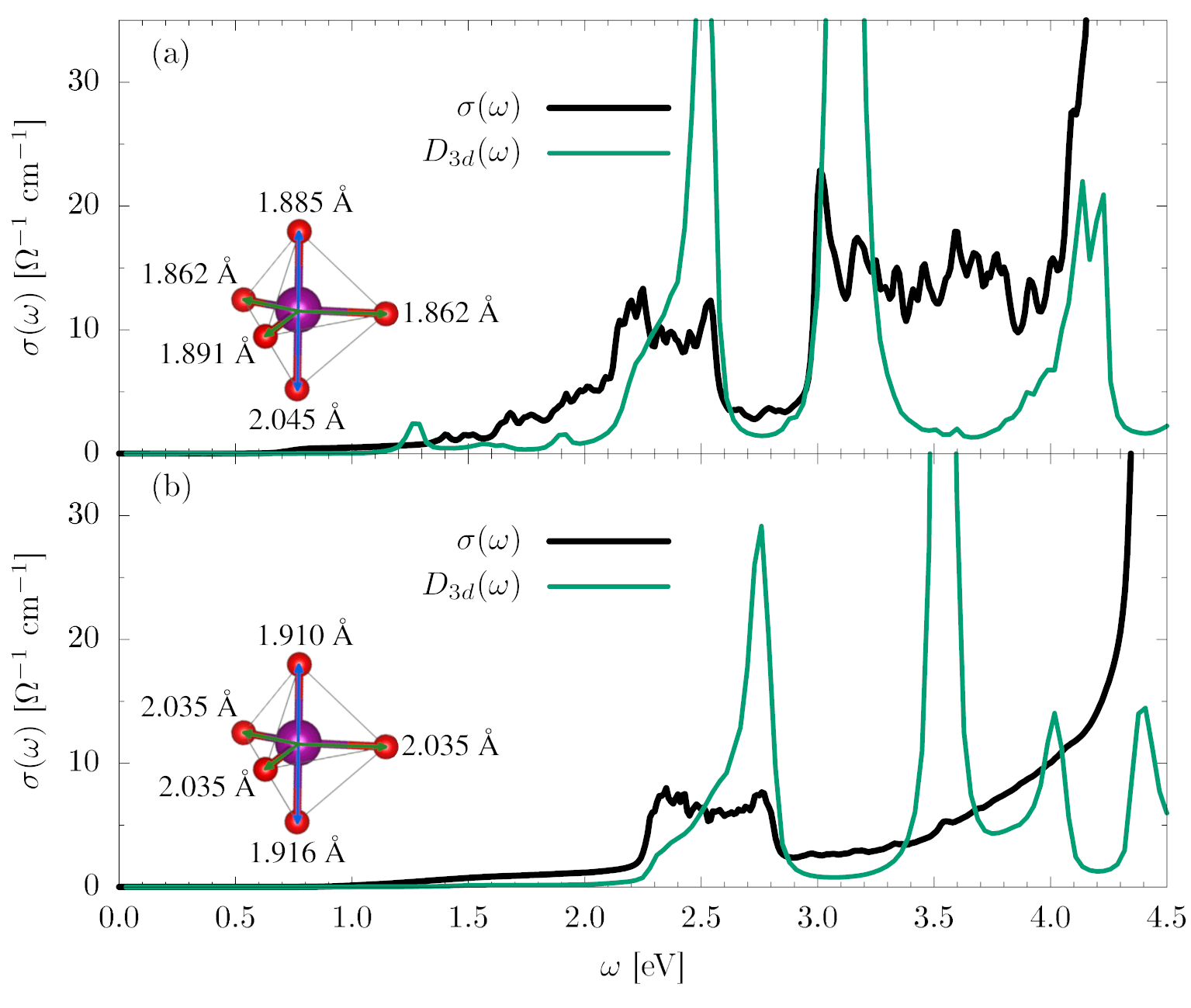}
	\caption{\textbf{Effect of asymmetric axial Mn-O bonds.} \textbf{(a)}   The optical conductivity (black) of YIn$_{0.92}$Mn$_{0.08}$O$_3$ with asymmetric axial Mn-O bond lengths shows a characteristic two-peak structure between \unit[1.2-4]{eV} arising mainly from Mn $d-d$ optical transitions (detail of \Cref{fig:bs_dos_opt}c). \textbf{(b)} A simulation with symmetric axial Mn-O bonds shows that the second peak centred around \unit[3.5]{eV} in the partial Mn 3$d$ JDOS (green) disappears in the optical conductivity (black) due to the influence of optical transition matrix elements. Input electronic structures were computed with mBJ@DFT+DMFT.}
	\label{fig:oc_bl_comp}
\end{figure}

\begin{figure}[t!]
	\includegraphics[width=\columnwidth]{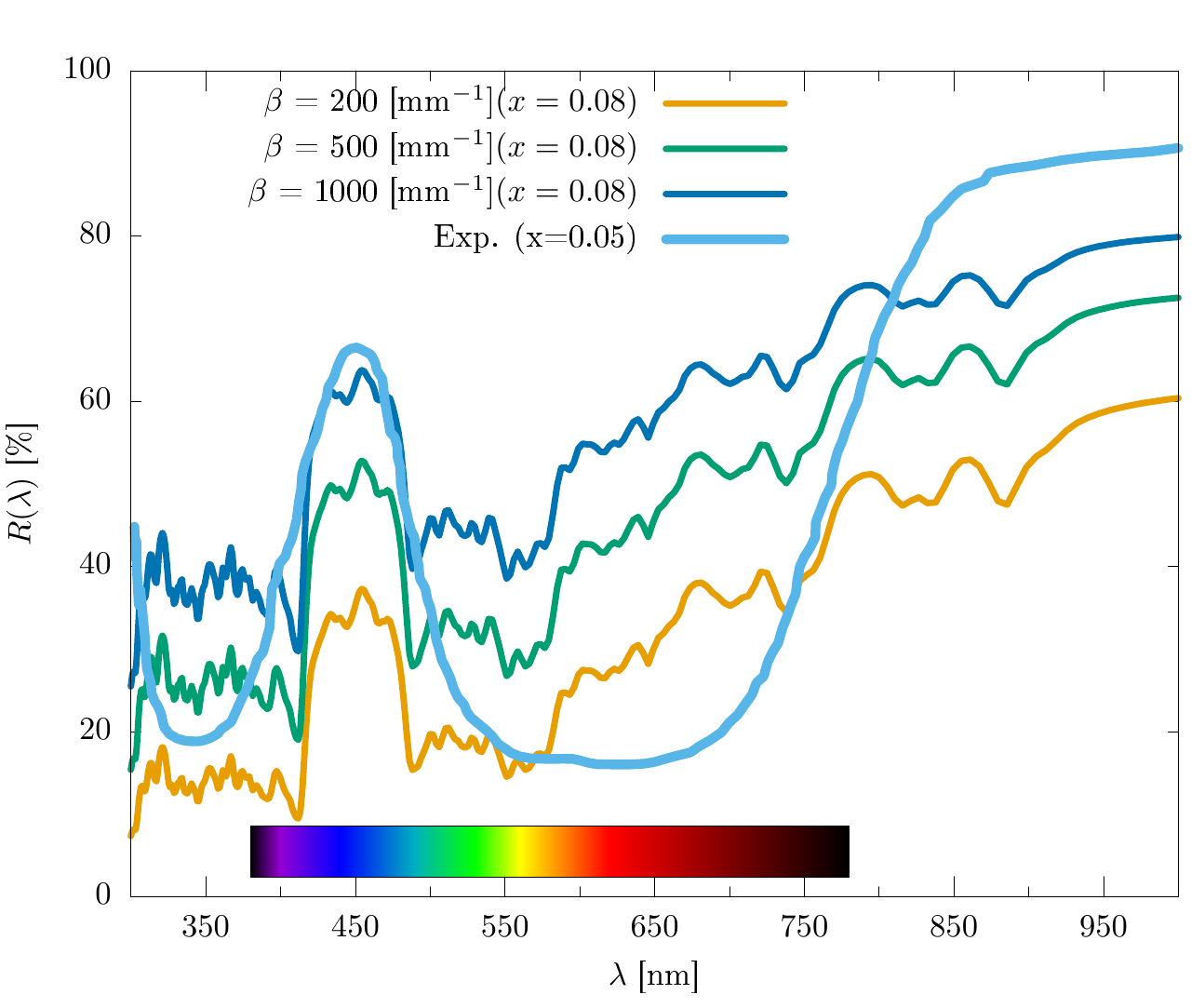}
	\caption{\textbf{Computed diffuse reflectance $R(\lambda)$ compared to experiment.} The theoretical curves for various values of the scattering parameter $\beta=\unit[200,500,1000]{mm^{-1}}$ (in orange, green and dark blue, respectively) all show a prominent reflectance peak in the blue region, which agrees well with experiment (light blue, data from Ref.~\onlinecite{YInMnTiZn_purple}, for $x=\unit[5]{\%}$). In the red region of the spectrum, the computed diffuse reflectance is too high compared to experiment.}
	\label{fig:Rdiff}
\end{figure}

In order to identify the effect of the axially distorted TBP on the optical response of YIn$_{0.92}$Mn$_{0.08}$O$_3$, we perform calculations with symmetric axial Mn-O bonds for comparison, using the crystal structure relaxed within LSDA+$U$. The latter did not show any asymmetry in the axial bonds (see Appendix~\ref{app1}, Table~\ref{tab:large}).
\Cref{fig:oc_bl_comp} represents a direct comparison between the optical conductivity $\sigma(\omega)$ of YIn$_{0.92}$Mn$_{0.08}$O$_3$ calculated with (a) asymmetric axial Mn-O bonds as in \Cref{fig:bs_dos_opt} and (b) assuming symmetric axial Mn-O bond lengths in the TBP. While \Cref{fig:oc_bl_comp}a shows the characteristic two-peak structure of $\sigma(\omega)$ already discussed in \Cref{fig:bs_dos_opt}c,  \Cref{fig:oc_bl_comp}b looks strikingly different. The partial Mn 3$d$ JDOS, $D_{3d}(\omega)$, still displays two peaks centred at \unit[2.7]{} and \unit[3.5]{eV}, respectively. Compared to \Cref{fig:oc_bl_comp}a, these peaks are located at slightly different energies due to the change in the local crystalline environment of the Mn$^{3+}$ ions.
More importantly, the main difference between $\sigma(\omega)$ in \Cref{fig:oc_bl_comp}a and b is the absence of the second peak at \unit[3.5]{eV} for symmetric TBP.
Since the principle difference between the JDOS $D_{3d}(\omega)$ and the optical conductivity $\sigma(\omega)$ stems from the transition-matrix elements $v_{\alpha\vb{k}}$ (see \Cref{eq:cond}), we assign the suppression of the peak at \unit[3.5]{eV} to the latter. Intuitively, one can indeed suspect the axial distortion---introduced by the asymmetric TBP---to further relax the optical dipole selection rules. 
This conjecture is supported by inspecting the momentum-resolved spectra (compare \Cref{fig:bs_dos_opt}a with \Cref{fig:bands_symm} of Appendix \ref{app3}): In the case of the undistorted TBP---with symmetric axial Mn-O bonds---the Mn-states are visibly less dispersive and hence closer to the atomic limit, for which the dipole selection rule applies.
The correspondence between dispersion and transition-matrix elements is explicit in the Peierls approximation, in which inter-unitcell transitions are weighted with $v_{\alpha\vb{k}}=\hbar^{-1}d/dk_\alpha H_0(\vb{k})$, where $H_0$ is the non-interacting Hamiltonian.\cite{PhysRevB.80.085117}

In the experimental investigations of Ref.~\onlinecite{Sarma2018} 
it had already been suspected that the axial distortion of the TBP is important for the blue coloration of YIn$_{1-x}$Mn$_{x}$O$_3$.
Our optical simulations fortify this understanding: We identify
the structure-driven change in transition-matrix elements as the essential requirement:
Only asymmetric axial Mn-O bonds yield the two-peak structure in the visible-range optical absorption
necessary for the blue color.

\subsection{Diffuse reflectance and color of the pigment}

\begin{figure}[t!]
	\includegraphics[width=\columnwidth]{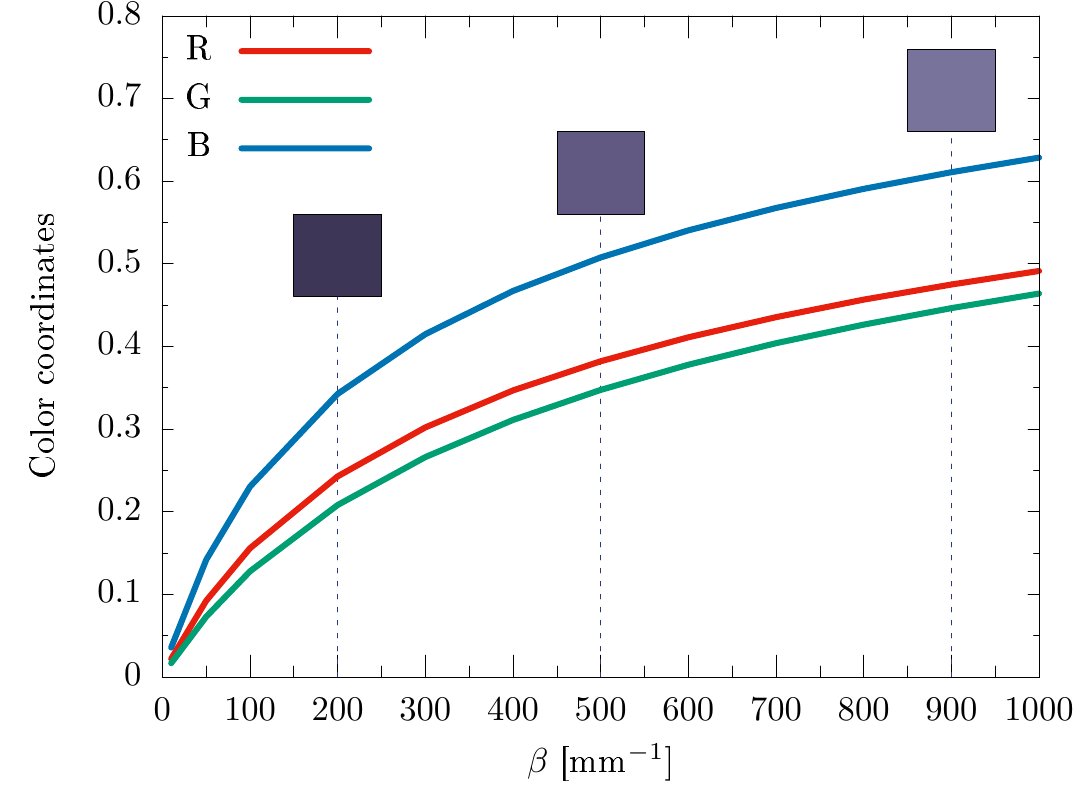}
	\caption{\textbf{Simulated color coordinates} of YIn$_{0.92}$Mn$_{0.08}$O$_3$, $R,G,B\in [0,1]$, calculated as a function of the scattering parameter $\beta$ for a non-diluted sample ($c_\%=1$) and apparent color (squares).}
	\label{fig:cc}
\end{figure}

We continue by computing the diffuse reflectance $R(\lambda)$ 
as a function of wavelength $\lambda=2\pi c/\omega$, using \Cref{eq:rdiff} and the optical conductivity from mBJ@DFT+DMFT. 
 In \Cref{fig:Rdiff} we compare the theoretical diffuse reflectance of YIn$_{0.92}$Mn$_{0.08}$O$_3$, for several values of the scattering parameter $\beta$,
 to the experiment of Ref.~\onlinecite{YInMnTiZn_purple}. We notice that the theoretical curves all display a pronounced reflectance peak in the blue region, which nicely matches with experiment. This reflectance peak arises from the discussed two-peak structure in the optical conductivity and is responsible for the blue color.
 With increasing $\beta$ also $R(\lambda)$ increases, 
 as more incoming light is back-scattered, while the overall shape of $R(\lambda)$ does not change significantly. On the low-energy side of the visible spectrum,  the agreement between theory and experiment is less convincing. Our simulations show a non-negligible refectance especially in the red region, which can be traced back to the weak onset of optical transitions in the \unit[1.2-2.2]{eV} energy range.

Finally, we compute the apparent color of the pigment, assuming a sample of concentration $c_\%=1$.
\Cref{fig:cc} shows the resulting $(R, G, B) \in [0,1]^3$ color coordinates as a function of $\beta$ with $(0,0,0)$ and $(1,1,1)$ being black and white, respectively. 
 Since the reflectance depends on $\beta$, also the color coordinates $(R, G, B)$ do.
 Our results show that the blue $B$ component dominates over the entire $\beta$-range, as expected.
Sizable admixtures from the $R$ and $G$ components result in an overall steal-blue color in the simulation, while experimental probes exhibit a deeper blue coloration.

\section{Conclusions}
We have presented the first attempt to compute the optical response and color of the blue pigment YIn$_{1-x}$Mn$_x$O$_3$ from first principles. We employed the recently developed mBJ@DFT+DMFT approach and confirmed---after its application to the rare-earth fluorosulfide pigments\cite{Galler2021_LnSF}---that this approach represents a useful computational tool for addressing the electronic structure of correlated pigment materials. In YIn$_{1-x}$Mn$_x$O$_3$, and pure YInO$_3$, the semilocal mBJ exchange potential successfully corrects the band gap between O 2$p$ and In 5$s$ states, which is significantly underestimated in LDA. The DMFT, instead, is able to treat the strong on-site Coulomb interaction in the Mn 3$d$ shell. 

We found that the optical response of YIn$_{0.92}$Mn$_{0.08}$O$_3$ close to the blue part of the spectrum is strongly influenced by the quasi-atomic Mn $3d$ states: Optical $d$-$d$ transitions yield a well-defined two-peak structure in the absorption, necessary for the blue coloration. The splitting of the Mn 3$d$ states is driven by multiplet effects and the local crystalline environment around the Mn$^{3+}$ ions. An important finding of our work is that asymmetric axial
distortions of the trigonal bipyramid surrounding the manganese impurities are a prerequisite for the blue color:
They 
weaken the atomic dipole selection rule, so that a pronounced two-peak structure can appear in the $d-d$ absorption.
Our work thus establishes a direct link between the asymmetry of the Mn$^{3+}$  coordination polyhedron and the blue coloration of YIn$_{0.92}$Mn$_{0.08}$O$_3$.
The weak absorption in the red-yellow region, due to optical transitions from Mn 3$d$ to the dispersive bottom of the In 5$s$ dominated conduction band, makes the pigment appear steal-blue in our calculations. 
Future investigations need to address the effect of Mn-disorder,
as well as trends for varying Mn-concentration.

\begin{acknowledgments}
 This work is based on the results of the Master thesis \cite{mastersthesis_VR} of V.\ R.\ at TU Wien. We thank Leonid V.\ Pourovskii, James Boust and Silke Biermann for helpful discussions. We further acknowledge financial support by Schr\"odinger Fellowship J-4267 of the Austrian Science Fund (FWF). Calculations were performed on the Vienna Scientific Cluster VSC4.
 \end{acknowledgments}

\appendix
\label{appendix}
\section{Electronic structure of YInO$_3$}\label{app1}
\begin{figure}
	\centering
	\includegraphics[width=\columnwidth]{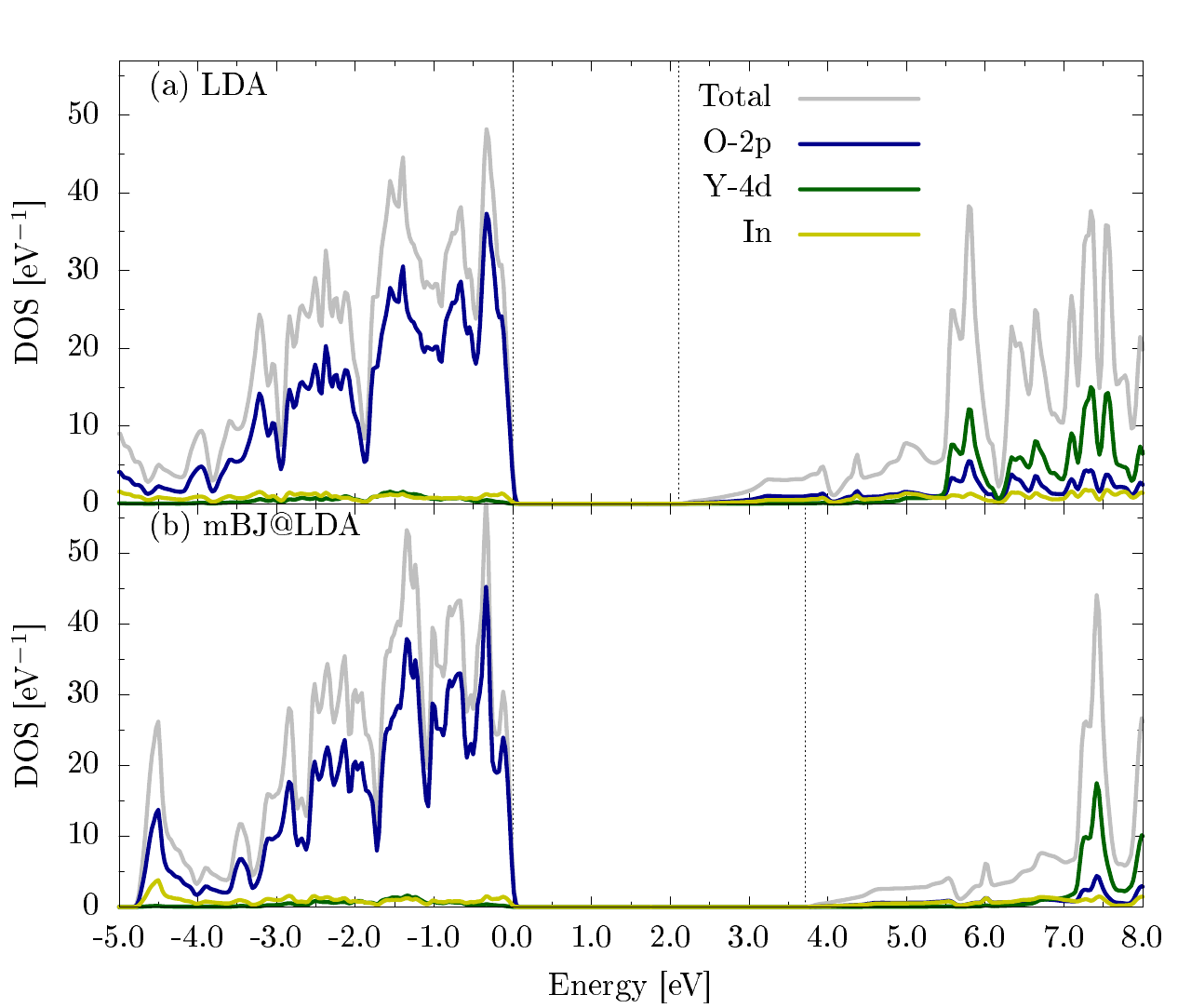}
	\caption{Density of states of YInO$_3$ calculated with \textbf{(a)} LDA and \textbf{(b)} mBJ@LDA. The dashed vertical lines indicate the band gap.}
	\label{fig:mbj}
\end{figure}

In \Cref{fig:mbj} we show the density of states (DOS) of pure YInO$_3$ computed within \textbf{(a)} LDA and \textbf{(b)} mBJ@LDA. The latter refers to the perturbative use\cite{hong_mbj_2013} of the mBJ potential\cite{tran_mbj_original} on top of a converged LDA calculation. In LDA, the band gap between the O 2$p$ dominated valence band and the bottom of the conduction band of mainly In 5$s$ character is only \unit[2.1]{eV} wide and thus significantly underestimated compared to its experimental value of approximately \unit[3.8]{eV} (which can be inferred from the diffuse reflectance of YInO$_3$ measured in Ref.~\onlinecite{Subramanian2009}). The mBJ potential, which effectively mimics non-local exchange, corrects the band gap to \unit[3.71]{eV}, which is in good agreement with experiment. Since such a perturbative use of the mBJ potential gives good results for pure YInO$_3$, we conclude that it is also well suited for YIn$_{1-x}$Mn$_x$O$_3$, with small Mn concentrations $x$.

\section{Summary of computed bond lengths}\label{app2}
In \Cref{tab:large} we summarise the results of our  structural relaxations performed within LDA and LSDA+$U$. Reported are computed axial and equatorial $M$-O bond lengths $\ell$ for pure YInO$_3$ ($x=0$), YIn$_{1-x}$Mn$_x$O$_3$ ($x=0.083$) and YMnO$_3$ ($x=1$). For comparison, we also reproduce experimental values  for $x=0,0.05,1$ extracted from Ref.~\onlinecite{Sarma2018}. For small Mn concentrations $x=0.05,0.083$, experiment and LDA relaxations show a clear axial asymmetry, in that one of the axial Mn-O bonds in the TBPs expands by $\sim10\%$.  Equatorial bond lengths slightly differ from experiment. Remarkably, the LSDA+$U$ relaxation for $x=0.083$ does not show any axial distortion of the TBPs. This difference might originate from the artificial magnetic order assumed in LSDA+$U$. However, we cannot exclude that our simulations converged to a local energy minimum often encountered in LSDA+$U$ calculations.
\begin{table*} 
	\bgroup
	\def\arraystretch{1.5}
	\begin{tabular}{c||c|c||c||c|c||c|c|c}
		&  \multicolumn{2}{c||}{x=0} & x= 5\% & \multicolumn{2}{c||}{x= 8.3\% } & \multicolumn{3}{c}{x = 1} \\
		\hline
		bond & \shortstack{$\ell$ [\AA]\\Exp.} & \shortstack{$\ell$ [\AA] \\ LDA} & \shortstack{$\ell$ [\AA]\\Exp.} & \shortstack{$\ell$ [\AA] \\ LDA} & \shortstack{$\ell$ [\AA] \\ LSDA+$U$} & \shortstack{$\ell$ [\AA]\\Exp.} & \shortstack{$\ell$ [\AA] \\ LDA} & \shortstack{$\ell$ [\AA]\\ LSDA+$U$} \\
		\hline
		\hline
		axial 1 & 2.089 & 2.088  & 1.879 & 1.891 & 1.910  & 1.848 & 1.856 & 1.867 \\
		\hline
		axial 2 & 2.093 &  2.111 & 2.087 & 2.045 & 1.916   & 1.882 & 1.875 & 1.869\\
		\hline
		equatorial 1 &  2.097 & 2.130 & 1.976 & 1.862 & 2.035 & 1.966 & 2.163 & 2.066\\
		\hline
		equatorial 2 & 2.127 & 2.122 & 1.976 & 1.862 & 2.035 & 2.118 & 2.002 &2.058\\
		\hline
		equatorial 3 & 2.127  & 2.122 & 2.087 & 1.885 & 2.035 & 2.118 & 2.002 &2.058\\
		\hline
	\end{tabular}
	\caption{$M$–O bond lengths $\ell$ of  YIn$_{1-x}$Mn$_x$O$_3$, for various Mn concentrations $x$ (for $x=0$, In-O bonds are reported; for all other $x$ Mn-O ones). Shown are the results of the structural relaxations within LDA  and LSDA+$U$ ($U=\unit[5]{eV}$, $J=\unit[0.5]{eV}$), as well as experimental data taken from Ref.~\onlinecite{Sarma2018}.} \label{tab:large}
	\egroup
\end{table*}

\section{Spectral function with symmetric axial Mn-O bonds}\label{app3}
\begin{figure}[t]
	\includegraphics[width=0.5\textwidth]{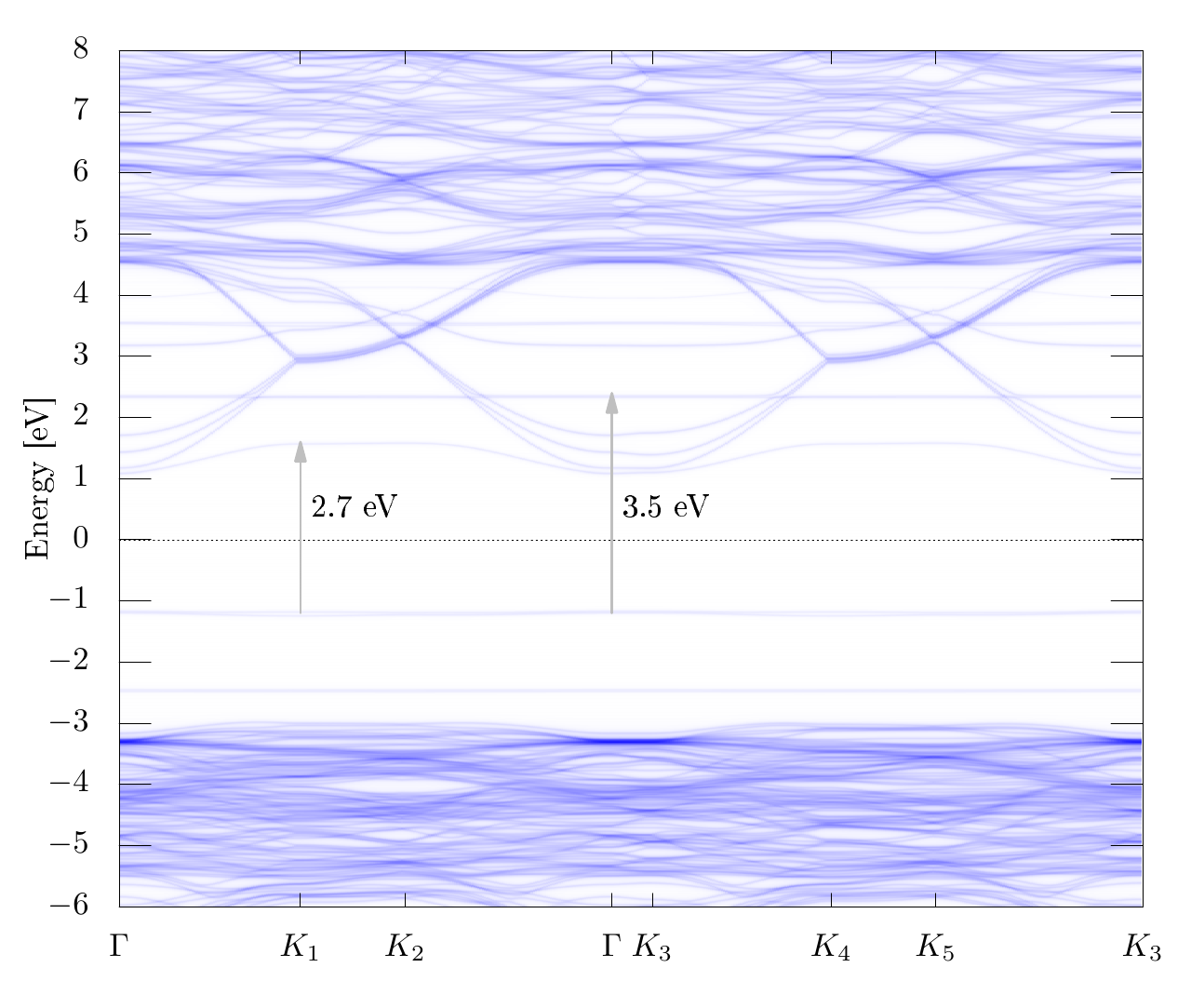}
	\caption{$\vb{k}$-resolved spectral function of YIn$_{0.92}$Mn$_{0.08}$O$_3$ with symmetric axial Mn-O bonds (bond lengths from LSDA+$U$ relaxation, see Table~\ref{tab:large}). The Mn 3$d$ states are flatter and more atomic-like than in  \fref{fig:bs_dos_opt}a.}
	\label{fig:bands_symm}
\end{figure}

\Cref{fig:bands_symm} shows the \textbf{k}-resolved mBJ@DFT+DMFT spectral function of YIn$_{0.92}$Mn$_{0.08}$O$_3$ assuming symmetric axial Mn-O bonds in the TBPs (structure relaxed within LSDA+$U$, bond lengths as specified in \Cref{tab:large}). The Mn 3$d$ bands are flatter and more atomic-like compared to \Cref{fig:bs_dos_opt}, completely suppressing the $d-d$ transitions in the optical conductivity at \unit[3.5]{eV}, see \fref{fig:oc_bl_comp}(b).

\begin{figure}
	\centering
	\includegraphics[width=0.8\columnwidth]{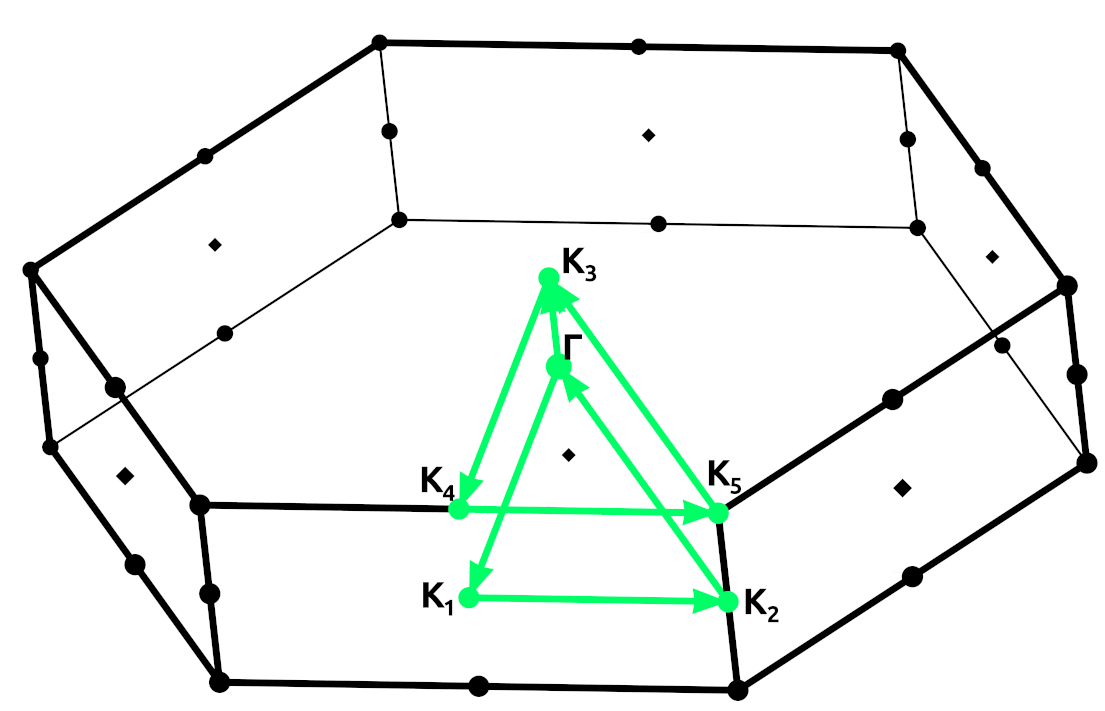}
	\caption{Primitive Brillouin zone (BZ) of YIn$_{0.92}$Mn$_{0.08}$O$_3$ plotted with \texttt{XCrySDen}.\cite{xcrysden} The $\vb{k}$-path through the BZ ($\Gamma - K_1 - K_2 - \Gamma - K_3 - K_4 - K_5 - K_3$) is marked by green arrows.	
	}
	\label{fig:bz_x8}
\end{figure}

The \textbf{k}-path through the hexagonal Brillouin zone (BZ) of YIn$_{0.92}$Mn$_{0.08}$O$_3$, employed in \Cref{fig:bands_symm} and \Cref{fig:bs_dos_opt}a,  is shown in \Cref{fig:bz_x8}.    

%

\end{document}